\documentclass[10pt]{article}

\usepackage{graphicx}
\usepackage[affil-it]{authblk}
\usepackage{amsmath}
\usepackage[T1]{fontenc}
\usepackage[utf8]{inputenc}
\usepackage{etoolbox}
\usepackage{lmodern}
\usepackage{caption}
\usepackage{setspace}
\usepackage{hyperref}
\usepackage{geometry}
\usepackage{titlesec}
\titleformat{\subsection}[runin]{}{}{}{}[]
\titleformat*{\section}{\LARGE\bfseries}
\titleformat*{\subsection}{\large\bfseries}

\usepackage{lineno}
 
\begin{document}

\title{\bf Phase resetting and intermittent control at critical edge of stability \\ as major mechanisms of fractality in human gait cycle variability}

\author[a]{Chunjiang Fu}
\author[a]{Yasuyuki Suzuki} 
\author[b]{Pietro Morasso}
\author[a]{Taishin Nomura \thanks{Corresponding author: taishin$@$bpe.es.osaka-u.ac.jp}}

\affil[a]{Graduate School of Engineering Science, Osaka University, Osaka 5608531, Japan}
\affil[b]{Center for Human Technologies, Istituto Italiano di Tecnologia, 16152 Genoa, Italy}

\renewcommand\Authands{ and }
\makeatletter
\patchcmd{\@maketitle}{\LARGE \@title}{\fontsize{12}{14.4}\selectfont\@title}{}{}
\makeatother

\renewcommand\Authfont{\fontsize{11}{14.4}\selectfont}
\renewcommand\Affilfont{\fontsize{9}{10.8}\itshape}


\date{}

\maketitle

\begin{abstract}
The fractality of human gait, namely, the long-range correlation that characterizes scale-free fluctuations of gait descriptors, such as the stride intervals during steady-state walking, depends on the well-tuned organization of the sensorimotor controller. Gait fractality is apparent in healthy young adults but tends to disappear in elderly individuals and neurological patients. Therefore, its partial loss may be indicative of pathological conditions. Despite its potential to be used as a dynamical biomarker for fall risk assessment, the mechanistic origin of gait fractality has been investigated by only a few studies, and even less attention has been devoted to the link between gait fractality and gait stability. Here, we propose a novel computational model of gait by addressing the flexibility-stability trade-off first, and then by showing that gait fractality is a natural consequence of the developed control mechanisms, including phase resetting and intermittent control, which supplement instability in a linear feedback controller operated at stability's edge. The results suggest that pathological gait, characterized by joint-rigidity and/or loss of fractality, may be caused by dysfunction in some of these mechanisms.
\end{abstract}


\vspace{0.5 cm}

Gait cycle fluctuation during steady-state human walking, examined typically by a variation in a series of stride intervals between time instants of consecutive ground-contact events for either foot, exhibits a long-range correlation that decays in a power-law manner \cite{Hausdorff2007555}. This variability is known as {\it gait fractality}, described with the astonishing fact that a given interval is influenced statistically by the long-term history over thousands of past intervals, as in a class of scale-free physiological phenomena including beat-to-beat heart rate variability \cite{PhysRevLett.70.1343, Goldberger19022002} and behavioral activity \cite{Gu2320}. Gait fractality is characterized by the rate of power-law decay, referred to as {\it the scaling exponent}, which involves critical information for quantifying pathological and age-related alterations in the gait control system \cite{Pailhous1992181, Hausdorff349, Frenkel-Toledo2005,DINGWELL2010348}: the exponent is close to unity, corresponding to $1/f$ fluctuation in healthy young adults, and it tends to be 1/2, corresponding to uncorrelated white noise in elderly and patients with neurological diseases including Huntington's and Parkinson's diseases \cite{Hausdorff1997aging, Hausdorff2009cha}. Moreover, the exponent could discriminate fallers and non-fallers among patients with high-level gait disorder \cite{HERMAN2005178}. 

Despite the high expectations of using the scaling exponent of gait fractality as a dynamical biomarker for the risk of falls \cite{Hausdorff2007555}, only a few studies have investigated the underlying sensorimotor mechanisms that may be responsible for fractal gait \cite{Hausdorff349, Ashkenazy2002662, Ahn2013, Dingwell201014}. Moreover, there are even less studies that relate gait fractality directly to gait stability and the risk of falls \cite{Bruijn20120999}. Considering the two decades of growing researches suggesting the usefulness of gait fractality, a theoretical account is keenly needed for gait fractality to become an established principle, in addition to being an acknowledged index in human gait research. 

A key concept here for understanding the deep rationale of the mechanisms leading to fractality of behavior is to take into account the general principle that too little variability is not necessarily favorable for physiological systems, as they could imply rigidity of the system, whereas too much variability may indicate instability of the system \cite{Stergiou2006,Todorov200911478}. That is, the human motor control system might be designed to avoid rigidity and instability, while aiming to achieve flexibility and stability simultaneously. However, there is a competition between these two motor features, leading to a trade-off, because flexibility and stability of movements are often characterized in an opposite manner as regards joint impedance: flexibility is indeed associated with small change-rates of restoring joint torque against perturbations in joint angle and angular velocity from the equilibrium state whereas stability implies high impedance levels \cite{Hogan1103644,Burdet2001}. The mechanisms that can resolve the flexibility-stability trade-off to achieve balanced functional performance levels are not trivial at all. 

We showed in previous studies that joint flexibility and postural stability can be achieved simultaneously in the investigation of human quiet standing \cite{Bottaro2008473,Asai2009,Suzuki201255} and steady-state bipedal gait \cite{Yamasaki2003, Yamasaki2003221, Fu20140958}; in particular, we demonstrated that types of time-discontinuous, nonlinear neural feedback controllers play a crucial role in stabilizing the body dynamics, while maintaining flexibility of the joints. For both postural and gait control, the key computational module is a mechanism called {\it the intermittent control} that exploits stable modes of the saddle-type unstable equilibrium of the passive dynamics of the body without active feedback control: the unstable upright equilibrium for standing and the unstable limit cycle oscillation for walking. 

In this study, we focuses on the gait model considering a multi-link rigid body system that is actuated by a feedforward (FF) controller and a conventional time-continuous small-gain proportional-derivative (PD) feedback controller for tracking a prescribed periodic desired joint angle profile, with the addition of well-timed brief activations of time-discontinuous feedback controllers. In particular, two types of discontinuous feedback controllers have been proposed previously. One is {\it an intermittent controller} (INT), as mentioned above, that brings the state vector of the walking body close to the stable manifold of the unstable limit cycle to be stabilized during brief ``on-periods'' of the controller, and then leaves the state point to flow along the stable manifold toward the unstable limit cycle during ``off-periods'' of the controller \cite{Fu20140958}. The other is {\it a phase resetting controller} (PR) that compensates timing of fluctuations of foot-contact-events due to endogenous and/or exogenous perturbations, in which the phase of the desired periodic joint-angle trajectory is advanced or delayed impulsively in response to an occurrence of every early or late event, respectively \cite{Schillings2093,Yamasaki2003, Yamasaki2003221}. As we show in the results section, by combining the three types of feedback controllers (PD, INT, and PR), we can achieve at the same time flexibility and stability. Joint flexibility in the model is obtained by operating with small gains of the continuous PD controller that maintain joint impedance at appropriate low levels. By tuning the PD controller in such a manner, we would induce instability of the gait, but this danger is avoided in a robust way by the subtle action of the two time-discontinuous controllers (INT and PR) that are activated briefly only at specific timings of the gait cycle: at double-stance phases for the intermittent controller \cite{Fu20140958} and immediately after heel-contact-events for the phase resetting controller \cite{Yamasaki2003, Yamasaki2003221}. 

A deeper motivation of the study was the expection that the mechanisms responsible for resolving the flexibility-stability trade-off might lead naturally to gait fractality, since joint flexibility is a basis of variability, provided that gait stability can be achieved even with such small joint impedance. Based on this working hypothesis, we performed dynamic simulations of the gait model with endogenous torque noise. In order to elucidate possible mechanisms necessary for the gait fractality, we systematically analyzed the relationship between gait stability and fractality in the model with different combinations of the four types of neural gait controllers defined above, and determined the conditions for the genesis of fractal gait.

\section*{Results}
We considered four combinations of four types of gait controllers to stabilize a limit cycle solution representing a periodic gait of the mechanical model with seven rigid-links ({\it Methods}), along with a set of periodic desired joint angle profiles prescribed for hip, knee, and ankle joints using motion-captured gait data \cite{Fu20140958,Yamasaki2003}. A FF controller ({\it Methods}) generating joint torques necessary for steady-state walking and a conventional PD feedback controller ({\it Methods}) at knee, ankle and hip joints to track the desired profiles with the feedback gains of $P=(P_k, P_a, P_h)$ and $D=(D_k, D_a, D_h)$ were always used for every combination. FF+PD controllers provide a basis of the gait, with which the steady-state gait can be attained stably if $P$-gains are large enough with a certain amount of $D$-gains.\footnote{$D=(10, 10, 10)$ Nm$\cdot$s/rad was used throughout the study.}

\subsection*{FF+PD controllers.}
Fig. 1A exemplifies a simulated time series of the stride intervals for the model with FF+PD controllers, when endogenous noisy torques of white Gaussian were additively superposed on the joint torques, exhibiting frequent stride-to-stride alternans between increase and decrease in the intervals. As shown in the magnified time series with noise and a sample of transient dynamics with a slow decay for the same model without noise (corresponding to the impulse response of the model), the alternans were an inherent property of the model\footnote{The alternans is associated with the fact that the dominant mode characterized by Floquet multipliers of the limit cycle of the model with FF+PD is a pair of complex conjugates with an amplitude close to unity and an argument greater than $\pi/2$ (see Fig. 4 of \cite{Fu20140958}), inducing period-2-like short period oscillations. Note that a real-valued Floquet multiplier near -1 (with the argument of $\pi$) generates a perfect alternans, which was not the case for this model.}. As the short period alternans with the slow decay in the stride intervals results in the anti-persistent correlation, the scaling exponent $\alpha$ estimated as the slope of linear regression for the plot of the detrended fluctuation analysis (DFA for quantifying scale-free long-range correlations in fractional Brownian motions; {\it Methods}) was almost zero (Fig. 1A-right). 

The anti-persistency in gait cycle variability shown in Fig. 1A was common across all values of $P$-gain within the stability regions (the light/dark-gray-colored regions in Fig. 2A-1st-row), as indicated by the corresponding color-coded values of the scaling exponent $\alpha$ in Fig. 2B-1st-row for the model with FF+PD. That is, the gait mechanics stabilized by the conventional linear PD feedback controller could never exhibit the fractal gait regardless of the $P$-gain values. Note that the $P$-gain used for Fig. 1A is indicated by the arrow-heads in the 1st rows of Figs. 2A and 2B.

\begin{figure}[tbhp]
\centering
\includegraphics[width=0.6 \linewidth]{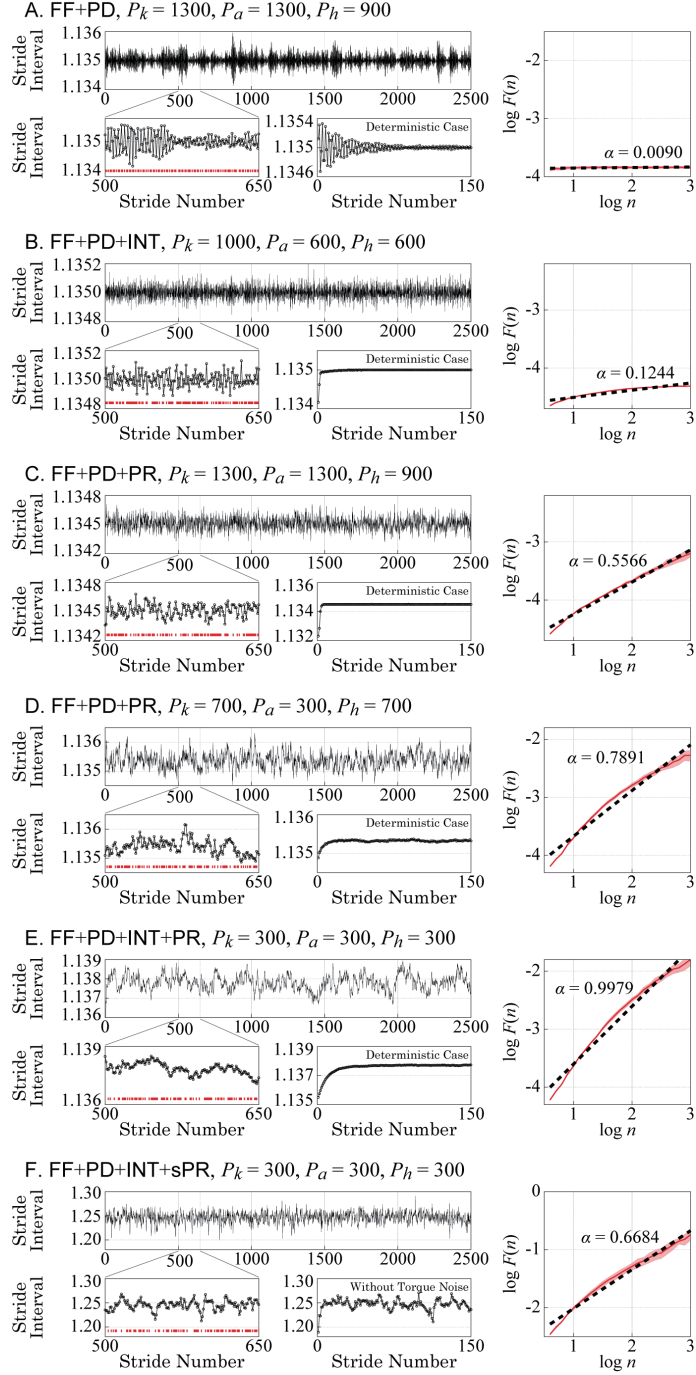}
\caption{Stride interval time series of the gait model actuated by different combinations of feedback controllers with and without endogenous white Gaussian torque noise ($\sigma=0.003$ Nm) and graphs of DFA. (A) FF+PD controllers for $P=(1300,1300,900)$ Nm/rad. (B) FF+PD+INT controllers for $P=(1000,600,600)$. (C) FF+PD+PR for $P=(1300,1300,900)$. (D) FF+PD+PR $P=(700,300,700)$. (E) FF+PD+INT+PR for $P=(300,300,300)$. (F) The same as (E), but a stochasticity was introduced in the amount of phase resetting. For each case, the small panel located left-below the long-run series of 2,500 strides is the magnification of the series for 150 strides, in which red bars beneath the waveform indicate the intervals showing the increase-decrease-increase or the decrease-increase-decrease alternations in the stride intervals. Densely distributed red bars imply anti-persistent behaviors of the stride intervals. The small panel located right-below represents the deterministic transient dynamics of the model without noise from an initial condition located slightly away from the steady-state limit cycle. The right-hand panel is a plot of DFA with the fluctuation size $F(n)$ against the scales from $10^{0.6}$ to $10^{3}$ stride number $n$ for the ten long-run trials, wherein the thickness of the red curve represents the standard deviations of the ten estimates of the $F(n)$. The scaling exponent value of $\alpha$ was estimated from the linear regression (dashed line).}
\label{fig:fig1}
\end{figure}

\subsection*{FF+PD+INT controllers.}
The use of the INT controller ({\it Methods}) in addition to FF+FD broadened the stability region substantially (Fig. 2A-2nd-row), as shown in \cite{Fu20140958}. However, it did not contribute to the emergence of fractal gait (Fig. 1B and Fig. 2B-2nd-row). Indeed, the short-period alternans in the stride intervals was still present\footnote{The mechanism of short-period alternans in the model with FF+PD+INT is not the same as that in the model with FF+PD, which can be seen from the non-oscillatory deterministic impulse response. For the $P$-gains smaller than those used for the model with FF+PD, the limit cycle of the model with FF+PD (as the model with FF+PD+INT during "off-period" of INT) is unstable, for which the pair of complex conjugate multipliers existed for the model with large gain FF+PD moves outside the unit circle and located typically near or on the real axis in the right-half plane (see Fig. 4 of \cite{Fu20140958}), associated with an unstable manifold of the limit cycle, along which the model during "off-period" of the INT controller falls away from the limit cycle. The alternans here is due to the random positioning of the state point on either side of the unstable manifold that is separated by the stable manifold at every timing of the onset of "on-period" of the INT controller.} as in the model with FF+PD, leaving the correlation in the stride intervals anti-persistent with the scaling exponent $\alpha$ close to zero.

\begin{figure*} 
\centering
\includegraphics[width=1.0\linewidth]{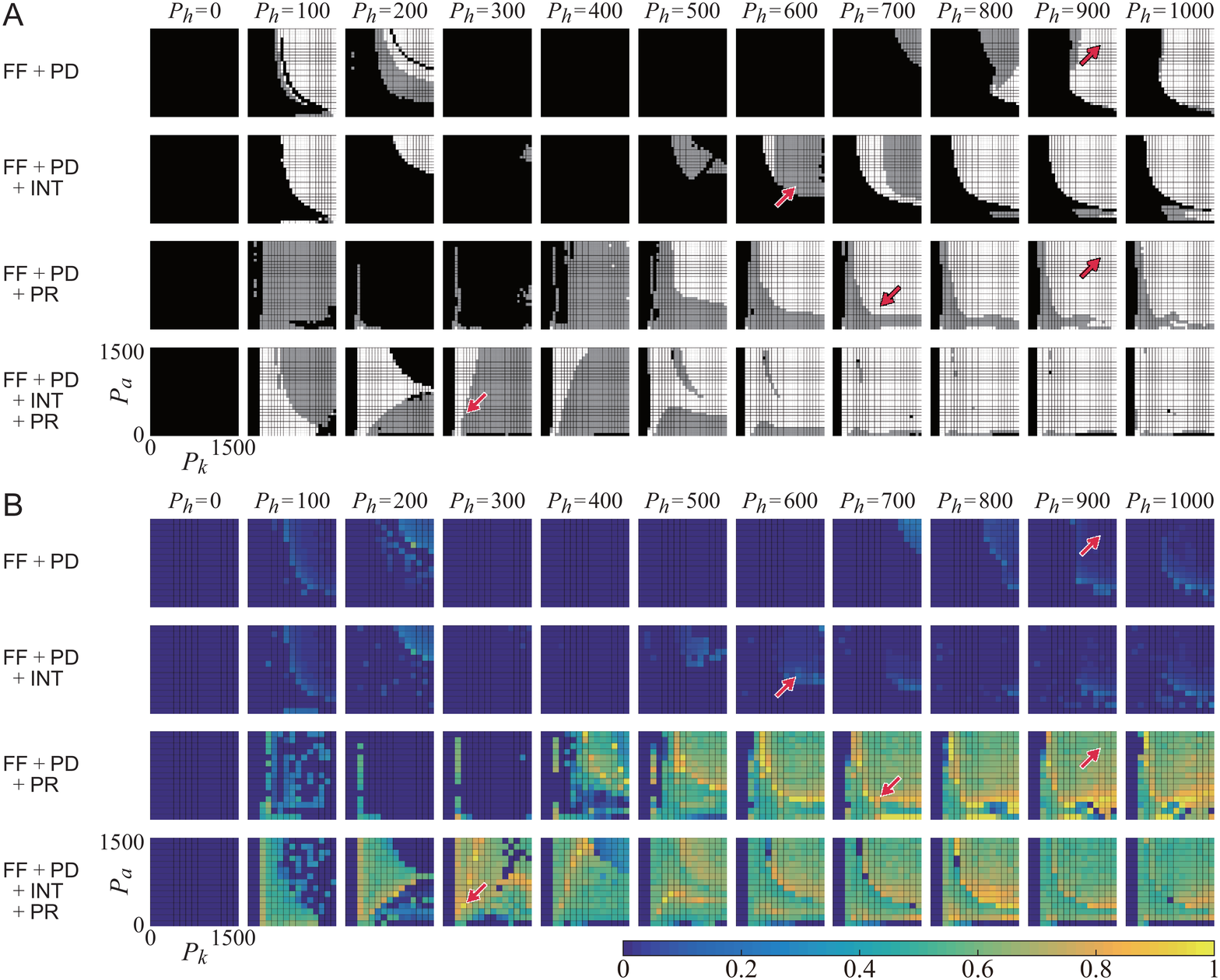}
\caption{Stability and scaling exponents as the functions of the proportional gains ($P$-gains) of the time-continuous PD feedback controller for four combinations of four neural controllers (FF, PD, INT, PR). All $D$-gains were fixed at 10 Nm$\cdot$s/rad. (A) Stability regions are indicated in light and dark gray in the $P$-gain parameter space in each panel, which means that the gait model could not walk stably in black regions. The panels aligned in each row represent the $P_k$-$P_a$ planes for 11 different $P_h$ values of 0, 100, $\cdots$, 1000 Nm/rad. Any stable limit cycle in the light gray region was close enough to the limit cycle $\gamma$ defined by Eq. (\ref{eq:ODE_FF}) representing the prescribed desired periodic gait, whereas the limit cycles in the dark gray regions were differ substantially from the desired trajectory. Thus, the border between light and gray regions in each panel, if exists, represents a bifurcation set at which the stability of the limit cycle altered and a new limit cycle appeared. Stability of limit cycle (either $\gamma$ or the one bifurcated from $\gamma$) was lost completely at another border between (light/dark) gray and black regions. Both types of borders represent the critical edges of stability. (B) Scaling exponent $\alpha$ as the functions of $P$-gain values, where values of the exponents between 0 and 1 are color-coded defined at the bottom of the panels. All simulations for (B) were performed with the endogenous white Gaussian torque noise of $\sigma=0.003$ Nm. For both (A) and (B), the four rows of panels were obtained for the model with different combinations of the controllers. 1st-row: FF+PD,  2nd-row: FF+PD+INT. 3rd-row: FF+PD+PR. 4th-row: FF+PD+INT+PR. The arrow-heads in the 1st, 2nd, 3rd, and 4th rows indicate the $P$-gain values used for the simulations shown in Figs. 1A, 1B, 1C, 1D and 1E(and also F), respectively. $1/f$-like fluctuation was observed at the grids with orange-yellow colors, which were distributed near the critical edges of stability.}
\label{fig:fig2}
\end{figure*}

\subsection*{FF+PD+PR controllers.}
The stride interval fluctuation became white-noise-like when the PR controller ({\it Methods}) was appended to the FF+PD controllers (Fig. 1C), where the deterministic impulse response was non-oscillatory and the degrees of alternans in the stochastic dynamics were substantially reduced, leading to an exponent $\alpha$ close to 1/2 (Fig. 1C-right). Moreover, the use of PR controller broadened the stability region in the $P$-gain parameter space more effectively than INT (Fig. 2A-3rd-row) as expected by our previous studies \cite{Yamasaki2003, Yamasaki2003221}, although the geometry of the limit cycle was altered as the $P$-gain approaches the marginal area of the extended stability regions, which is indicated by the light and dark gray colors of the stability regions in Fig. 2A. That is, the limit cycle in the light gray regions oscillates according to the prescribed reference dynamics (defined as $\gamma$ in {\it Methods}), whereas the one in the dark gray regions is qualitatively different from the prescribed $\gamma$, referred to as $\gamma'$ for convenience in this sequel. The emergence of $\gamma'$ is due to destabilization of $\gamma$ in some dimensions, representing a bifurcation phenomenon, but the overall gait dynamics remained stable as long as the $P$-gain is located in the dark gray regions. In this way, the border between the light and gray regions represents the bifurcation curve or the bifurcation set in the $P$-gain parameter space.  

In the extended stability regions of the model with FF+PD+PR, the scaling exponent $\alpha$ values of about 1/2 were widely distributed as shown by the light-blue and green regions of Fig. 2B-3rd-row. Interestingly, the exponent exhibited large values between 0.8 and 1.0 for specific areas located along the bifurcation curve of the limit cycle within the light gray area of the $P$-gain plane in Fig. 2A-3rd-row (Fig. 1D and the orange-yellow arcs in Fig. 2B-3rd-row). That is, the long-range power-law correlation was observed when the $P$-gain for the model with FF+PD+PR was located at the critical edge of stability of the prescribed limit cycle $\gamma$.  

\subsection*{FF+PD+INT+PR controllers.} The combined use of INT and PR controllers dramatically broadened the stability region in the $P$-gain parameter space (Fig. 2A-4th-row), by which the model could establish stable walking for very small $P$-$D$ gains, implying stable gait with very flexible (compliant) joint dynamics. In particular, the stability regions with the prescribed $\gamma$ (the light gray regions) occupied larger areas than those for the model with FF+PD+PR, as compared with the light gray regions between the 3rd and 4th rows of Fig. 2A. As a result, the border between the light and dark gray regions marched into the small $P$-gain regime, wherein the exponent $\alpha$ values between 0.8 and 1.0 were widely distributed (the orange-yellow regions in Fig. 2B-4th-row) along the edge of stability (the bifurcation curve). Fig. 1E is such an example of stride interval dynamics showing the long-range correlated $1/f$-like fluctuation with noise and the deterministic impulse response with a slow decay approximated by a power-law function $n^{-1}$. 

Here, by reference to a clinical report showing a larger stochasticity in the amount of PR for non-Parkinsonian patients with a symptom of freezing of gait (FoG) than for Parkinsonian patients without FoG \cite{Tanahashi2013}, we examined how the $1/f$-like dynamics in the model with FF+PD+INT+PR could be influenced by a stochasticity introduced on the amount of PR ({\it Methods}). It was shown that the stochasticity of the PR could lead to a loss of the long-range correlation (Fig. 1F). This result suggests that the randomness of the PR is one of the possible mechanisms that cause a loss of gait fractality.

\section*{Discussion}
\subsection*{Summary.}
Based on the numerical simulations of the gait model with different types of controllers, we showed that the model with endogenous torque noise can exhibit the fractal gait, if the PR mechanism is incorporated as a feedback controller, and the gains of time-continuous PD feedback controller for tracking a reference gait motion are tuned at the critical edge of stability. Moreover, combined use of the INT feedback controller with PD and PR controllers remarkably lowers the critical gains of PD for stability and broadens the gain-parameter regions that exhibit gait fractality, by which the fractal gait becomes a commonplace phenomenon in terms of the area in the $P$-gain parameter space. Thus, these three conditions clarified in this study could be the major mechanisms responsible for the genesis of fractal gait. It is logically apparent that the gait dynamics that satisfied these conditions simultaneously naturally resolve the flexibility-stability trade-off. The results suggest that pathological gait, characterized by joint-rigidity and/or loss of fractality \cite{Hausdorff1997aging, Hausdorff2009cha}, may be caused by dysfunction in some of these mechanisms. 

\subsection*{Stability versus flexibility.}
A novel outcome of this study, besides the fractal issue, is that the combined use of PR and INT with FF+PD controllers could establish gait stability with extremely small overall joint impedance (the small $P$-gain values). For example, the gain values of $P_a=P_k=P_h=100$ Nm/rad and $D_a=D_k=D_h=10$ Nm$\cdot$s/rad corresponding to the parameter point that is located in the stability region of Fig. 2A-4th-row are quite comparable with experimentally estimated joint stiffness of 200-300 Nm/rad for the lower extremities during human walking \cite{Shamaei2013a, Shamaei2013b}. To the best of our knowledge, there might be no other human bipedal gait models that can establish a stable walking with such a small joint impedance that allows flexible joint dynamics.  

\subsection*{Limitations.} The gait model used here is simplified from many points of view. In particular, it describes only two-dimensional sagittal motion during walking, unlike others \cite{ROOS20102929}, and is limited to steady-state walking. The prescribed periodic desired joint-angle trajectory may have a limited physiological substance, though it can be conceptually considered as a set of motor output signals produced by central pattern generators \cite{Ijspeert2017} that exhibit PR in response to perturbations \cite{Schillings2093}. Because the desired joint-angle trajectory used for the model is fixed as prescribed, and it cannot be altered actively against perturbations, except its oscillating phase by the PR, the overall gait stability of the model is probably more limited than that in a healthy human subject. For this reason, the noise intensity used in this study could not be large enough, which made the variance of stride intervals smaller than experimental reality. In perspective, however, the limitations of the model could be overcome by another computational layer, in addition to the control layer investigated in this study, that addresses the human capability to adapt to changing environmental conditions by updating an original plan of action. 

\subsection*{Related studies.}
The mechanisms of stride-to-stride control that can generate the fractal gait is not trivial; thus, it is not easy to hypothesize possible mechanisms. However, such mechanisms might not be unique. Dingwell et al. have proposed a reconciling mechanism among {\it optimality} of control-signal (minimum intervention) and task-related error (errors in the walking velocity from a desired constant velocity), {\it redundancy} in stride length (distance) and stride interval (time) to achieve the desired constant walking speed (distance/time), and {\it stochasticity} by multiplicative motor noise, with a concept of goal equivalent manifold (GEM) associated with the redundancy, which can generate the long-range power-law correlation of stride interval fluctuation along the GEM \cite{Dingwell201014}. Unlike our neuromechanical modeling, their optimal feedback control model has no underlying mechanics, and thus cannot be directly related to stability (falls) of the mechanical body dynamics. That is, the GEM is defined in the abstract space concerning the control strategy. Nevertheless, the gait model studied here and the GEM-based optimal control model could be deeply inter-related with each other, as discussed in detail by \cite{Suzuki201255} for a model of postural fluctuation during standing, as the GEM, a kind of uncontrolled manifold defined for kinematic redundancy of the human body \cite{Latash200226}, might be associated with the stable manifold used for the INT control as shown in our previous study \cite{10.3389/fnhum.2016.00618}. That is, the redundancy for strategy and that for body kinematics should be closely inter-related.   

Emergence of fractal gait near the edge of stability in this study is reminiscent of a recent report showing that energetic costs are minimized by the control at stability's edge during expert stick balancing \cite{Milton2016}. Moreover, the best-fit model parameters for the stick balancing experiment using the INT control model in our previous study were also located at the critical edge of stability \cite{10.3389/fncom.2016.00034}. Although the mechanical energy consumptions in our model did not show significant differences among four examined combinations of controllers (results not shown), which was probably because the fluctuation of the gait dynamics of the model was too small to make any difference, criticality in stability and the long-range correlated fluctuation might generally be closely related. In this regard, clarification of the interrelationship among criticality, optimality, and fractality is the significant future challenge in motor control research. 

\subsection*{Risk of falls and adaptability.}
In view of importance of the fall risk assessment for prediction and prevention of falls in the aging and aged societies, the reliability of the gait index available so far needs to be improved substantially \cite{Verghese2009896}. Although some of the traditional measures are not necessarily capable of characterizing motor functions properly, newly developed measures, such as those quantifying complexity and/or nonlinear dynamics of fluctuation are expected to contribute to improving the quality of assessments \cite{Bruijn20120999, TOEBES2012527, Zhou2017}.  A difficult, but yet scientifically interesting aspect of gait stability has been demonstrated well by Bruijn et al \cite{BRUIJN20091506} who showed the absence of a consistent pattern of correlations between orbital stability of the gait limit cycle evaluated by the local Lyapunov exponent and the size (the standard deviation) of variability of gait dynamics. More importantly, alterations in a metric value can represent different meanings depending on the underlying mechanisms that alter the gait dynamics. For example, a recent study shows that increases in gait variability, which has been reported to be associated with fall history, may not imply impaired stride-to-stride control of walking in healthy older adults, but may be due to increased neuromotor noise \cite{DINGWELL2017131}. Gait fractality quantified by the scaling exponent may also face a similar difficulty for practical use as a dynamical biomarker. As suggested by the current study, the appropriate combination of feedback controllers can broaden the stability regions into the small-gain regime, which makes the fractal gait with flexible joint dynamics robust against changes in the gain-parameters. However, this fact {\it per se} does not imply high stability of gait, but rather adaptability of the gait control system. Although local dynamics analyzed by any methodologies, including gait fractality, can never characterize the global stability of gait, such as stability determined by a largeness of the basin of attraction of the limit cycle for example \cite{Yamasaki2003, Yamasaki2003221}, we can speculate that a highly adaptable gait control system is more capable of exhibiting high global stability. Thus, the fractal gait measure should contribute to the assessment fall risks, with other measures characterizing different aspects of gait dynamics.  


\section*{Methods}
The equations of motion of the gait model, the controllers of FF, PD, INT, PR, and stochastic PR,  and DFA are described below.  
\subsection*{Equations of motion of the model.}
The model with seven rigid-links in the sagittal plane from our previous work \cite{Fu20140958} was used in this study. The general coordinates of the model can be described as 
\begin{equation}
q=(q_1,q_2,\cdots,q_9)^T\equiv (\theta,x,y,\theta_a^l,\theta_k^l,\theta_h^l,\theta_a^r,\theta_k^r,\theta_h^r)^T, 
\end{equation}
where $\theta$, $x$ and $y$ are the tilt angle, horizontal position and vertical position of the head-arm-trunk (HAT) center of mass (CoM). $\theta_a^i$, $\theta_k^i$, and $\theta_h^i$ are the ankle, knee, and hip joint angles of the left ($i=l$) and right ($i=r$) limbs. The equation of motion is represented as 
\begin{equation}
J(q)\ddot{q}+B(q, \dot{q})+K(q) +G(q, \dot{q})= U,
\label{eq:motion_equation} 
\end{equation}
where $J(q)$, $B(q,\dot{q})$, $K(q)$ and $G(q, \dot{q})$ are the inertia matrix, centrifugal and Coriolis torque, gravitational torque and the ground reaction force, respectively. $U=(0,0,0,U_a^l,U_k^l,U_h^l,U_a^r,U_k^r,U_h^r)^T$ on the right-hand side represents the joint torques acting on the six joints. $G(q, \dot{q})$ was modeled using a nonlinear spring and damper acting forces on the heel and the toe of either foot \cite{Yamasaki2003, Fu20140958}. 

\subsection*{The desired joint angle profiles.}
Joint angles of the model are represented by one part of the coordinate $q$, and we denote it by 
\begin{equation}
\hat{q}\equiv (q_4,\cdots, q_9)^T. 
\end{equation}
Time profiles of the desired joint angles or the desired joint angle trajectories, denoted by  $\hat{q}^d(\phi)$ as a function of the gait phase $\phi$ or $\hat{q}^d(t)$ as a function of time $t$, are defined for $\hat{q}$ as
\begin{equation}
\hat{q}^d(\phi) \equiv (q_4^d(\phi),\cdots, q_9^d(\phi))^T
\end{equation}
where the gait phase $\phi$ takes a value in $[0, T_c]$ with $T_c=1.135$ (sec) being the gait period of the steady-state walking, or equivalently 
\begin{equation}
\hat{q}^d(t) \equiv (q_4^d(\mbox{mod}_{T_c}(t+\phi_0)),\cdots, q_9^d((\mbox{mod}_{T_c}(t+\phi_0))^T, 
\end{equation}
for an appropriate initial phase $\phi_0$ of the desired joint angle trajectory, where $\mbox{mod}_{T_c}$ takes a modulo $T_c$ of the argument. 
Note that $\phi$ and $t$ are related as $\phi=\phi(t)\equiv\mbox{mod}_{T_c}(t+\phi_0)$. 
$\hat{q}^d(\phi)$ or $\hat{q}^d(t)$ was obtained as the Fourier expanded series of a human walking motion-captured data as detailed in \cite{Yamasaki2003, Fu20140958}. 

\subsection*{FF controller.}
A benefit of the model used here with the desired joint angle trajectory $\hat{q}^d(\phi)$ is that the model can walk stably when the time courses of all of the six joints in Eq. (\ref{eq:motion_equation}) are kinematically constrained by $\hat{q}^d(\phi)$, by which only three kinematic degrees of freedom $(q_1, q_2, q_3)$ and their derivatives $(\dot{q}_1, \dot{q}_2, \dot{q}_3)$ are remained as dynamic variables \cite{Yamasaki2003}. A forward dynamic simulation of Eq. (\ref{eq:motion_equation}) with such kinematic constraints, after its transient dynamics, generates the corresponding time profiles of $(q_1^\text{ss}(\phi), q_2^\text{ss}(\phi), q_3^\text{ss}(\phi))$. In this way, we obtain a complete sequence of kinematics of the model during steady-state periodic gait as $\bar{q}(\phi)=(q_1^\text{ss}(\phi), q_2^\text{ss}(\phi), q_3^\text{ss}(\phi), q_4^d(\phi),\cdots, q_9^d(\phi))^T$, and the corresponding ground reaction force $G(\bar{q}(\phi),\dot{\bar{q}}(\phi))$. The inverse dynamics solution of Eq. (\ref{eq:motion_equation}) for the six joints, which are used as the outputs of the FF controller \cite{Fu20140958}, was obtained as
\begin{equation}
U_\text{ff}(\phi)=J(\bar{q}(\phi))\ddot{\bar{q}}(\phi)+B(\bar{q}(\phi), \dot{\bar{q}}(\phi))+K(\bar{q}(\phi)) +G(\bar{q}(\phi), \dot{\bar{q}}(\phi)).
\label{eq:U_ff} 
\end{equation}
The equation of motion with the FF controller is formulated as
\begin{equation}
J(q(t))\ddot{q}(t)+B(q(t), \dot{q}(t))+K(q(t)) +G(q(t), \dot{q}(t))= U_\text{ff}(\phi(t)). 
\label{eq:ODE_FF} 
\end{equation}
In this study, we considered the periodic trajectory of the state point defined as $\bar{x}(\phi)=(\bar{q}(\phi), \dot{\bar{q}}(\phi))$ in the 18-dimensional state space of Eq. (\ref{eq:ODE_FF}), referred to as $\gamma$ in this section. By definition, $\gamma$ is {\it always} a periodic solution of Eq. (\ref{eq:ODE_FF}), regardless of its stability, although the FF controller alone cannot stabilize $\gamma$. 

\subsection*{PD feedback controller.}
As in \cite{Fu20140958}, the PD controller generates the feedback torques defined as
\begin{equation}
U_\text{fb}(t) = P(\hat{q}^{d}(t+\phi_0)-\hat{q}(t))+D(\dot{\hat{q}}^{d}(t+\phi_0)-\dot{\hat{q}}(t)), 
\end{equation}
or equivalently
\begin{equation}
U_\text{fb}(\phi) = P(\hat{q}^{d}(\phi)-\hat{q}(\phi))+D(\dot{\hat{q}}^{d}(\phi)-\dot{\hat{q}}(\phi)),
\label{eq:U_fb} 
\end{equation}
where
\[
\begin{array}{lll}
P &=& \mbox{diag}(P_a,P_k,P_h,P_a,P_k,P_h), \\
D &=& \mbox{diag}(D_a,D_k,D_h,D_a,D_k,D_h). 
\end{array}
\]
The equation of motion with the FF+PD controller is formulated as
\begin{equation}
J(q)\ddot{q}+B(q, \dot{q})+K(q) +G(q, \dot{q})= U_\text{ff}(\phi)+U_\text{fb}(\phi). 
\label{eq:ODE_FFPD} 
\end{equation}
Because $\gamma$ is the solution of Eq. (\ref{eq:ODE_FF}), it is also the solution of Eq. (\ref{eq:ODE_FFPD}). FF+PD controller can stabilize the limit cycle $\gamma$, if $P$ and $D$ are large enough. In this study, $D$ was fixed at $D_a=D_k=D_h=10$ Nm$\cdot$s/rad. $P$-gain values were altered systematically in the range of  $P_h\in[0,1000]$ Nm/rad, $P_a\in[0,1500]$ Nm/rad, and $P_k\in[0,1500]$ Nm/rad. 

\subsection*{INT feedback controller.}
The INT feedback controller provides feedback torques only during an optimized brief time span, lasting for $w(=0.1)$ second, after the heel contact within every double support phase of the gait \cite{Fu20140958}. It is used together with FF+PD controllers, typically when $P$-$D$ gains of PD controller are small and located outside the stability regions. Because the INT controller is activated only in a sequence of brief time spans, it contributes to reducing the overall impedance necessary for maintaining the walk.

The INT controller exploits the stable manifold $W^{s}(\gamma)$ of unstable $\gamma$ as the solution of Eq. (\ref{eq:ODE_FFPD}) for small gains of PD controller, where $W^{s}(\gamma)$ was obtained locally around $\gamma$ using the monodoromy matrix (a linearized Poincar\'e map) with its stable eigenvalues (Floquet multipliers) and the corresponding eigenvectors. Let $\gamma(\phi)$ be a state point on $\gamma$ at phase $\phi$, which means $\gamma=\cup_{\phi}\gamma(\phi)$. Then, we can define $W^{s}(\gamma(\phi))$ as the intersection between $W^{s}(\gamma)$ and a Poincar\'e section of $\gamma$ passing through $\gamma(\phi)$ for the linearized Poincar\'e map. The INT controller, when it is activated, aims to force the state point of the model at the phase $\phi$ toward $W^{s}(\gamma(\phi))$, {\it not to} $\gamma(\phi)$, using the feedback torque defined as
\begin{equation}
U_\text{int}(\phi) = P^{+}(\hat{q}^{s}(\phi)-\hat{q})+D^{+}(\dot{\hat{q}}^{s}(\phi)-\dot{\hat{q}}), 
\label{eq:intermittent_controller} 
\end{equation} 
where $(\hat{q}^{s}(\phi),\dot{\hat{q}}^{s}(\phi))\in W^{s}(\gamma(\phi))$ with the gains of 
$P^{+}$ and $D^{+}$ defined as
\[
\begin{array}{lll}
P^{+} &=& \mbox{diag}(P_a^{+},P_k^{+},P_h^{+},P_a^{+},P_k^{+},P_h^{+}), \\
D^{+} &=& \mbox{diag}(D_a^{+},D_k^{+},D_h^{+},D_a^{+},D_k^{+},D_h^{+}). 
\end{array}
\]
$P_a^{+}$-$D_a^{+}$, $P_k^{+}$-$D_k^{+}$ and $P_h^{+}$-$D_h^{+}$ are the gains of the INT feedback controller acting on the ankle, knee and hip joints, respectively, of the left and right limbs. $U_\text{int}$ is supplemented to the right-hand-side of Eq. (\ref{eq:ODE_FFPD}) intermittently only during on-period defined as $[\phi_\text{on},\phi_\text{on}+w]$ for $w$ seconds. That is, the equation of motion 
with FF+PD+INT controllers can be described as 
\begin{equation}
J(q)\ddot{q}+H(q, \dot{q}) = U_\text{ff}(\phi)+U_\text{fb}(\phi) + U_\text{int}(\phi),
\end{equation}
if $\phi\in[\phi_\text{on-L},\phi_\text{on-L}+w]$ or $\phi\in[\phi_\text{on-R},\phi_\text{on-R}+w]$, which is during the on-period of INT controller, and  
\begin{equation}
J(q)\ddot{q}+H(q, \dot{q}) = U_\text{ff}(\phi)+U_\text{fb}(\phi), 
\end{equation}
otherwise during the off-period of INT controller, where $H(q,\dot{q})\equiv B(q, \dot{q})+K(q) +G(q, \dot{q})$. In this study, $\phi_\text{on-L}$ and $\phi_\text{on-R}$ were set to the timings of 1.5 ms after every heel-contact event of the left and right feet, respectively. See \cite{Fu20140958} for details. 

\subsection*{PR controller with FF+PD.}
The PR controller used in this study resets the phase $\phi$ of the desired joint angle trajectory impulsively at every heel-contact of either foot. During steady-state walking of the model along the prescribed $\gamma$, heel-contact events of the left (right) foot occur periodically with the gait period $T_c$, for which the phases of left (right) heel-contact events are fixed at $\phi_L$ ($\phi_R$). During transient gait or in the presence of noise that perturbs the state point from $\gamma$, the phase of every heel-contact may deviate from the reference phase, i.e., $\phi_L$ or $\phi_R$. The PR controller shifts the phase $\phi$ of the desired joint angle trajectory to compensate such deviations. That is, when a heel-contact of the left (right) foot is detected, the phase $\phi$ of the desired trajectory is reset to $\phi_{L}$ ($\phi_{R}$). Thus, the amount of phase resetting is $\Delta(\phi)=\phi-\phi_L$ for a left heel-contact, and it is $\Delta(\phi)=\phi-\phi_R$ for a right heel-contact. If a heel-contact occurs earlier (later) than the reference timing, $\Delta(\phi)<0$ ($\Delta(\phi)>0$), by which the gait phase is advanced (delayed) from $\phi$ to $\phi-\Delta(\phi)$, which is equal to either $\phi_L$ or $\phi_R$ for both of advanced and delayed resettings. Note that the FF torques are also reset by the same amount when the desired joint angle trajectory is reset. It has been shown that the PR mechanism can increase gait stability and largely reduce the impedance necessary for stability \cite{Yamasaki2003221,Aoi2010}. 

Stochasticity in the amount of PR was introduced using a truncated normal distribution. That is, for every heel-contact event at phase $\phi$, the deterministic amount of PR $\Delta(\phi)$ was randomly modified as $\Delta(\phi)+\xi$ where $\xi\sim N(0,\sigma_r)$ with $\sigma_r$ was set to 10 ms and any outcomes beyond $2\sigma_r$ were discarded so that $n$ took values within $[-2\sigma_r, +2\sigma_r]$ only. 

\subsection*{FF+PD+INT+PR.}
When the PR and INT controllers were used together with FF+PD, the sequence of executing two time-discontinuous control mechanisms was detecting a heel-contact first, then resetting the phase of the desired joint trajectory and the FF torques. Then, the INT controller was turned on 1.5 ms after the PR was performed.

\subsection*{Torque noise.}
Stochastic fluctuation of the gait was induced by adding a set of six independent series of white Gaussian torque noise (with an identical standard deviation $\sigma=0.003$ Nm) acting on the 6 joints. Because the gait model in this study focused only on the recovery against small perturbations when the desired joint-angle trajectory and thus the geometry of the limit cycle were fixed, the stride interval fluctuation could not be large enough in comparison with the real human data. However, we intended to use large noise amplitude as much as possible. 

\subsection*{DFA.}
The standard procedure of DFA \cite{PhysRevE.49.1685,Peng199582} was applied for the stride interval time series data $\{x(i)\}_{i=0}^{N-1}$ of length $N$. We obtained an integrated time series $y(k) = \sum_{i=0}^{k-1}{(x(i) - \bar{x})}$, where $\bar{x}$ is the sample mean of the series. Then $\{y(k)\}_{k=1}^{N}$ was divided into equal segments of length $n$ without overlapping, so the segment number was $M=N/n$. In each segment of length $n$ with  an index $l$ ($l=1,\cdots,M$), a least square polynomial $p_n^{(l)}(k)$ as the trend of the segment was fitted to the data. Square of deviations in the segment $l$ were summed up to obtain $f_n^{(l)}=\sum_{k=(l-1)n+1}^{ln}{(y(k)-p_n^{(l)}(k))^2}$. The root of average square deviation of all segments was then calculated as $F(n)=[M^{-1}\sum_{l=1}^{M}f_n^{(l)}]^{1/2}$. For each $n$, the corresponding $F(n)$ was calculated, finally the scaling exponent $\alpha$ was obtained by the linear fitting of $\log F(n)$ and $\log n$. 

Longer data length is preferred to obtain reliable results. Typically in the literature, 1 hour walk including about 3,000 strides are tested experimentally \cite{Hausdorff19961448}. Here we adopt the higher standard of 3,000 strides for every $P$-gain parameter to quantify a scaling exponent. The orders of the polynomial ranging from 1 to 3 were examined for many preliminary sets of $P$-gain, for each of which we obtained 10 simulated trials of 3,000 strides to calculate a mean of 10 exponent values. We found that the exponent was not altered quantitatively for the different orders. Thus, we used the linear fitting (the order 1) for calculating the exponent for the wide-range parameter study, in which the exponent was calculated from only one trial of 3,000 strides for each $P$-gain due to significant computational time necessary for the simulations. 

\subsection*{Numerical simulation.}
Numerical simulations of the model were performed by integrating equations of motion simply using the Euler-Maruyama method \cite{doi:10.1137/S0036144500378302} with a time step of $\Delta t=10^{-5}$ seconds. The robustness of this numerical simulation method was tested by time step ten times larger and smaller. For each simulation, an initial state $(q(0),\dot{q}(0))$ and an initial phase $\phi_0$ were specified, in which the initial state was set close to the limit cycle such that $|q(0)-\bar{q}(\phi_0)|<\epsilon$ was satisfied for a given $\phi_0$. $\epsilon$ was set as $10^{-4}$ rad on the HAT tilt angle.

In the model with PR and INT controllers in the presence of noise, we considered that a heel contacted on the ground if the ground reaction force remained non-zero continuously more than 20 time steps. PR was performed immediately after a detection of heel-contact. Then, with a latency (during which only FF+PD controllers were employed) of $w=1.5$ ms (150 time steps), the INT controller was turned on, which lasted for $w = 0.1$ sec. 

All simulations were performed on workstation features 2 Intel Xeon E5-2687W (10 cores each at 3.10 GHz), with a 64 GB RAM. The operation system was CentOS 6.5, with Intel compiler 13.1 and job scheduler Lava 1.0.6. Typically, one trial of parameter scanning ($P_a$, $P_k$, and $P_h$ were varied between 0 and 1500 Nm/s with fixed $D$-gains at 10 Nms/rad) for the model with INT and PR controllers took about 10 days. 


\subsection*{Acknowledgements.}
This work was supported in part by JSPS grants-in-aid 26242041 and 26750147, MEXT KAKENHI Grant Number 16H01614 (Non-linear Neuro-oscillology), and MEXT grant for Post-K project (primary issue 2).

\subsection*{Author contributions.} T.N. conceived and designed the study. C.F. and Y.S. performed the modeling, simulations and data analysis.  C.F, Y.S., P.M., and T.N. wrote the paper.

\subsection*{Competing interests.} 
The authors have declared that no competing interests exist.


\end{document}